# Toxicological Evaluation of Phytochemicals and Heavy Metals in *Ficus exasperata* Vahl (Sandpaper) Leaves obtained in Birnin Kebbi, Nigeria

Tajudeen O. Yahaya[1a], Titilola F. Salisu[2], Israel O. Obaroh[3], Muyiwa A. Adelabu[4], Abdulrazaq Izuafa[1b], Jamilu B. Danjuma[5], Hamdalat Sheu[1c], Ibrahim B. Abdulgafar[1d]

[1]Department of Biological Sciences, Federal University Birnin Kebbi, PMB 1157, Kebbi State, Nigeria. E-mail: yahayatajudeen@gmail.com[a], izuafa.abdulrazaq@fubk.edu.ng[b], sheuhamdalat1198@gmail.com[c], abdulgafarbala2018@gmail.com[d]
[2]Department of Zoology and Environmental Biology, Olabisi Onabanjo University Ago-Iwoye, PMB 2002, Ago-Iwoye, Ogun State, Nigeria. E-mail: titilola.salisu@oouagoiwoye.edu.ng
[3]Department of Animal and Environmental Biology, Kebbi State University of Science and Technology, PMB 1144, Aliero, Nigeria. E-mail: obarohio@gmail.com
[4]Department of Cell Biology and Genetics, University of Lagos, PMB 101017, Akoka, Lagos, Nigeria. E-mail: adelabdiagnostics@gmail.com
[5]Department of Biochemistry and Molecular Biology, Federal University Birnin Kebbi, PMB 1157, Kebbi State, Nigeria. E-mail: jamiludanjuma11@gmail.com

Corresponding Author: Tajudeen Yahaya, E-mail: yahayatajudeen@gmail.com and yahaya.tajudeen@fubk.edu.ng

**Abstract**

**Objective:** *Ficus exasperata* Vahl (Sandpaper tree) is extensively used in Nigeria to treat diseases, but a dearth of documentation about its toxicity exists. This information is crucial because pollutants can contaminate medicinal plants. This study determined the heavy metal and phytochemical content of methanolic leaf extract of *F. exasperata* obtained in Birnin Kebbi, Nigeria.

**Material and Methods:** The lethality of the plant was also assessed using 70 wild shrimps divided equally into seven groups. Group 1 (negative control), groups 2 and 3 (positive controls) were exposed to 500 and 1000 ppm of formaldehyde, respectively; and groups 4-7 were exposed to 1000, 2000, 4000, and 8000 ppm of extracts, respectively, for 96 hours.

**Results:** The phytochemistry revealed high levels of flavonoids and saponins and moderate levels of tannins and phenols. The heavy metal analysis revealed non-tolerable levels of cadmium, copper, and lead, while zinc was within the tolerable limit. The negative control recorded 10% mortality, 1000 and 2000 ppm (20% each), 4000 ppm (70%), and 8000 ppm (100%).

**Conclusion:** These results inferred safe doses of the plant's extract in low and medium concentrations but toxic and fatal at high doses over a period of time. Consumers are advised to seek an expert's guidance before using it.

**How to cite**: Yahaya TO, Salisu TF, Obaroh IO, Adelabu M, Izuafa A, Danjuma JB, Sheu H, Abdulgafar IB. Toxicological Evaluation of Phytochemicals and Heavy metals in *Ficus exasperata* Vahl (Sandpaper) Leaves obtained in Birnin Kebbi, Nigeria. Plant Biotechnology Persa 2022; 4(2): 1-7.

## Introduction

Plants have been used to treat ailments since the dawn of time. As far back as the time of the earliest cavemen, archaeologists have discovered evidence of plants being used for medical purposes in practically every discovery. Herbal medicine is a part of the human heritage that spans cultures. Herbs have been utilized in traditional medicine in Europe, Asia, and India since the 1800s. Naturopathy, homeopathy, and other holistic treatments were extensively adopted by American doctors in the early 1900s to manage or treat certain diseases [1]. However, herbal medicine practices faded as modern medicine and pharmaceutical drugs advanced. But currently, there is a renaissance in plant-based medicines due to the adverse effects and high cost of synthetic medications. People are turning back to plant medicine as an alternative to synthetic medications because it represents safety [2]. About 90% of the world population currently uses herbal medicines to treat diseases [3]. Changes in lifestyles that favor natural products over synthetic ones may explain why people in developed countries are interested in herbal medicine again [3].

*Ficus exasperata* Vahl, also called sandpaper tree, is an important plant commonly used in plant medicine. It belongs to the genus Ficus and contains about 850 species of woody trees, shrubs, vines, epiphytes, and hemi-epiphytes [4]. *F. exasperata* is a rapidly growing tree that is domiciled in the rainforest, savannah, and beside rivers and streams [5]. *F. exasperata* is a small to medium-sized fig tree belonging to the banyan family, with a height of 20–30 meters (66–98 feet). To anchor itself in the earth and sustain hefty branches, the trunk develops aerial and buttressing roots [6]. *F. exasperata* contains a wide variety of bioactive compounds such as alkaloids, phenols, flavonoids, saponins, sterols, terpenes, glycosides, tannins, carbohydrates, and proteins [7, 8]. The mentioned phytoconstituents and others could be responsible for the efficacy of *F. exasperata* in treating different diseases worldwide. In Nigeria, *Ficus species* are used to treat piles, tuberculosis, ulcers, hypertension, microbial infection, hyperlipidemia, as well as asthma, diarrhea, diabetes, stomach aches, and constipation [9, 10].

However, concerns are rife about the safety of many medicinal plants, including *F. exasperata*, as the form of medicine gains popularity. The bulk of herbal products on the market are not standardized to recognized active ingredients, and quality control methods are not always strictly followed [11]. There have been reported cases of toxicities and hospitalization following the consumption of some medicinal plants or plant products [12]. This suggests that, even though medicinal plants have been used for decades, phytoconstituent analysis and toxicity studies are still required to know their safe doses [13]. This can be achieved through phytochemical and heavy metal analyses of the plants as well as a brine shrimp lethality assay. The brine shrimp assay is rapid, inexpensive, simple (no special equipment or techniques required), and uses a large number of organisms for statistical validation and a small amount of test sample, as little as 2 mg [14]. This study, therefore, employed heavy metal and phytochemical analyses as well as a modified brine shrimp lethality assay to determine the toxicity of *F. exasperata* leaves obtained in Birnin Kebbi, Nigeria. The plant is often used for medicine in Birnin Kebbi and environs. However, there is a dearth of information on the toxicity of *F. exasperata* in the city. Conduction of plant toxicities in each locality is important considering that pollutants are now widespread in the environment, which may potentially accumulate in plants and animals, all based on anthropogenic activities and soil geology.

## Materials and Method
### Collection of plant samples and preparation of extract

Fresh leaves of *F. exasperata* were harvested from the plant in Birnin Kebbi metropolis in November 2021 and identified by a taxonomist at the Department of Biological Sciences, Federal University Birnin Kebbi, Nigeria. A sample of the authenticated material with voucher no. FUBK123 was retained in the herbarium of the department. After that, the leaves were gently washed to eliminate contaminants and air-dried for one week in the shade. A laboratory mill (manufactured by Norris Limited, Poole, England) was used to grind the dry leaves into powder. Prior to usage, the ground plant material was kept in a desiccator. Fifty grams (50 g) of *F. exasperata* powder were dissolved in 500 ml of 95% methanol and left for 72 hours. The macerated methanolic extract was filtered through a muslin cloth, and then it was evaporated at 40 °C to a constant dry weight.

### Qualitative and quantitative phytochemical screening

A stock solution of the extract was prepared by dissolving 0.06 g in distilled water in a 50-mL beaker and making it up to the meniscus with distilled water. The stock solution was tested for the presence of alkaloids, tannins, flavonoids, saponins, terpinoids, phenols, quinone and cardiac glycosides as conducted by Yahaya et al. [12]. The phytochemicals detected in abundance in the extract of *F. exasperata* were thereafter quantified.

## Heavy metal analysis

The levels of copper (Cu), lead (Pb), cadmium (Cd), and zinc (Zn) in *F. exasperata* extract were determined following the procedures of Yahaya and Okpuzor [15]. One (1) g of the extract was placed in a clean beaker (100 ml) containing an analytical grade of 25 mL of aqua-regia and 5 mL of 30% $H_2O_2$. The mixture was digested at 80 °C until it became a homogeneous solution. After cooling, the solution was filtered into a 50-mL beaker and filled to the meniscus with deionized water. A UNICAM atomic absorption spectrophotometer (model 969) was used to measure the levels of the selected heavy metals.

## Brine shrimp lethality assay

The brine shrimp lethality assay was conducted as explained by Olowa and Nuñeza [38] but with some modifications. Seventy (70) shrimps were caught from Lagos lagoon and kept in the lagoon water in a well-ventilated animal house. A stock solution of the extract of *F. exasperata* was prepared by dissolving 13.8 g in 100 mL of distilled water. From the stock solution, 1000, 2000, 4000, and 8000 ppm of the extract were prepared. The shrimps were divided into 7 groups of 10 each. Group 1 was made a negative control and kept in the lagoon water; groups 2 and 3 were positive controls and exposed to 500 and 1000 ppm of formaldehyde, respectively; and groups 4-7 were exposed to 1000, 2000, 4000, and 8000 ppm of the extracts, respectively. Each group was checked for mortality after 4, 24, 48, 72, and 96 hours.

## Quality control and assurance

All the chemicals used were of high purity. All glass and plastic materials were washed and rinsed properly with distilled water and the reagent to be placed in. The accuracy of the heavy metal analysis was ensured by checking for background contamination of the samples in which blank samples were tested along with the test samples. Each heavy metal analysis was replicated thrice, and the level of reproducibility was ensured to be above 95%. As a result, the mean of the three values of each heavy metal was used for further analysis.

## Results
## Qualitative analysis of the plant

Table 1 reveals the phytochemicals detected in the extract of *F. exasperata* leaves obtained in Birnin Kebbi, Nigeria. Flavonoids and saponins were detected in abundance, tannins and phenols were detected in moderate amounts, cardiac glycosides and terpenoids were detected in trace quantities, and alkaloids were not detected.

**Table 1:** Phytochemicals detected in the extract of *Ficus exasperata* leaves obtained in Birnin Kebbi, Nigeria

| Phytochemicals | Inference |
|---|---|
| **Saponins** | +++ |
| **Flavonoids** | +++ |
| **Tannins** | ++ |
| **Phenol** | ++ |
| **Cardiac Glycosides** | + |
| **Terpenoids** | + |
| **Alkaloids** | - |

Note: - stands for absent; + indicates trace amount; ++ means moderate; +++ indicates abundance

## Quantitative analysis of the plant

The concentrations of flavonoids, saponins, tannins, and phenols, which were detected at moderate or abundant levels (as applicable) during qualitative analysis, are shown in Table 2. Flavonoids had the highest concentration, followed by saponins, phenols, and tannins, respectively.

**Table 2:** Mean concentrations of phytochemicals in the extracts of *Ficus exasperata* leaves obtained in Birnin Kebbi, Nigeria

| Quantifiable Phytochemicals | Concentration (mgml$^{-1}$) |
|---|---|
| **Flavonoids** | 0.45 ± 0.07 |
| **Saponins** | 0.41 ± 0.13 |
| **Phenols** | 0.27 ± 0.03 |
| **Tannins** | 0.15 ± 0.02 |

## Levels of heavy metals in the plant

The levels of Zn, Cd, Cu, and Pb in the extract of *F. exasperata* leaves are displayed in Table 3. With the exception of Zn, the heavy metals were present beyond the limits set for individual heavy metals by the World Health Organization (WHO) and Food and Agricultural Organization (FAO).

**Table 3:** Mean levels of heavy metals in the extract of *Ficus exasperata* leaves obtained in Birnin Kebbi, Nigeria

| Heavy metals | Levels (mgkg$^{-1}$) | Permissible limit [17] |
|---|---|---|
| **Zn** | 1.50 | 5.00 |
| **Cd** | 0.05 | 0.01 |
| **Cu** | 0.61 | 0.20 |
| **Pb** | 0.25 | 0.01 |

## Toxicity of the plant on shrimp

The toxicity of the extract of *F. exasperata* on the shrimps caught from Lagos lagoon was revealed in Table 4. The negative control group had a 10% mortality, while the groups that were exposed to 1000 and 2000 ppm of the extract each had a 20% mortality, the 4000 ppm group had a 70% mortality, and the 8000 ppm group recorded a 100% mortality.

**Table 4:** Mortality recorded by shrimps (n = 10 per group) exposed to different concentrations of leaf extract of *Ficus exasperata* obtained in Birnin Kebbi, Nigeria

| Group | Concentration (ppm) | Mortality (4hrs) | Mortality (24hrs) | Mortality (48hrs) | Mortality (72hrs) | Mortality (96hrs) | Percentage Mortality (%) |
|---|---|---|---|---|---|---|---|
| 1 | -ve | 0 | 0 | 0 | 0 | 1 | 10 |
| 2 | +v1 | 1 | 5 | 3 | 1 | - | 100 |
| 3 | +v2 | 1 | 9 | - | - | - | 100 |
| 4 | 100 | 1 | 1 | 0 | 0 | 0 | 20 |
| 5 | 200 | 0 | 1 | 0 | 0 | 1 | 20 |
| 6 | 400 | 0 | 2 | 1 | 0 | 4 | 70 |
| 7 | 800 | 0 | 3 | 3 | 0 | 4 | 100 |

Note: –ve = negative control group kept only in lagoon water; +ve 1 = positive control 1 exposed to 500 ppm of formaldehyde; +ve 2 = positive control 2 exposed to 1000 ppm of formaldehyde

## Discussion

This study determined the levels of phytochemicals and heavy metals in the extracts of *F. exasperata* leaves obtained in Birnin Kebbi, Nigeria, and also determined their toxicity using a modified brine shrimp lethality assay. *F. exasperata* is commonly used in traditional medicine in Nigeria, but there is a dearth of documented information about its toxicity. This study, therefore, aimed to determine the safe doses of the plant in order to prevent unintended fatalities among its users.

Tables 1 and 2 show that the plant is rich in phytochemicals, which justifies its efficacy in treating or managing several diseases mentioned earlier. However, the extract of the plant contained high levels of flavonoids and saponins, which may accumulate to toxic levels if the plant is taken constantly for a prolonged period of time. High doses of saponins may cause loss of appetite, diarrhea, malnutrition, and hepatic failure, among other health effects [18-20]. High levels of flavonoids can induce mild to moderate histopathological alterations in the liver and kidneys [21]. The results of the current study are consistent with almost all available studies on the plant. Notably, Kofie *et al*. [22] detected the mentioned phytochemicals in *F. exasperata* obtained from Kwame Nkrumah University of Science and Technology, Kumasi, Ghana. Ajayi *et al*. [23] also detected these phytochemicals in *F. exasperata* leaves obtained in Ilesha, Osun State, Nigeria. Nworu *et al*. [24] reported similar findings in *F. exasperata* obtained in Nsukka, Enugu State, Nigeria. However, the mentioned studies and the current study showed varying levels of the phytochemicals, which could be caused by several factors. According to Kumar *et al*. [25], temperatures and wind patterns (which vary worldwide) affect precipitation, which in turn affects plant architecture, flowering, fruiting, and phytochemical composition. The varied levels of the phytochemicals can also be caused by the different physiological and developmental stages of the *F. exasperata* used in each study, in which flowering plants produce more phytochemicals [26]. Environmental stress, mostly from pollutants like heavy metals and microorganisms, can also cause some phytochemicals to build up in plants [27].

Furthermore, Table 3 reveals that the plant extract contains non-permissible levels of Cd, Cu, and Pb, which again indicates that the plant can induce health hazards in consumers. In minute quantities, some heavy metals perform biological functions [28]. But at certain levels, they can build up in the body and deplete antioxidants, producing free radicals and other health risks [29]. The kidneys, liver, pancreas, lungs, bones, and testicles are the main targets of Cd toxicity [30]. Pb is neurotoxic and can cause mental retardation, particularly in children [30]. Pb and Cd can also increase the risk of cardio-vascular diseases [31]. Excessive Cu buildup in the body can cause respiratory and reproductive disorders [32]. The results of the present study are in line with those of Sunmola *et al*. [33], who detected non-permissible levels of Cd and Pb in the samples of *F. exasperata* obtained at Covenant University, Ota, Ogun State, Nigeria. Agrahari *et al*. [34] also detected non-permissible levels of Pb (the only heavy metal evaluated in the study) in the samples of *F. exasperata* obtained in Gorakhpur City, Uttar Pradesh, India. However, the results contrast with those of Tadesse *et al*. [35], who detected permissible levels of the majority of the heavy metals assessed in *F. exasperata* obtained in the Awash River Basin, Ethiopia. Ladipo and Doherty [36] also detected permissible levels of selected heavy metals in the samples of *F. exasperata* obtained in Mushin, Lagos, Nigeria. These inconsistencies could be due to varying heavy metal-emitting anthropogenic activities in the various localities in which the studies were conducted. It could also be caused by the varied geology of the localities. Because Birnin Kebbi (the study area of the current study) is not industrialized, the most probable source of these heavy metals in the plant is its natural deposits in the soil as well as agricultural inputs. Vehicular emissions and dust from mining activities could also be the sources of the heavy metals.

Table 4 indicates that 1000 and 2000 ppm doses of the extracts of the plant induced low mortality among the exposed shrimps when compared with the negative control. On the other hand, 4000 and 8000 ppm doses progressively induced high mortality, with 8000 ppm being more lethal and even causing the death of all the exposed shrimps at the end of the experiment. Furthermore, the results also revealed that a single high dose of the plant may not elicit side effects, as no death was recorded after a single high dose of the plant. But prolonged high dose consumption of the plant can induce toxic effects, as shown by the increasing mortality recorded by the shrimps with increasing duration of exposure. These findings suggest that low to medium doses of the plant extract in the range of 1000 to 2000 ppm may be safe for consumption, especially when heavy metal contamination is prevented. The result of the current study is consistent with that of Shemishere *et al*. [37], who reported non-lethality of a single low-to high dose of methanol extract of *F. exasperata* on some treated rats. But Shemishere *et al*. further observed that after repeatedly dosing the rats for 14 days, biomarkers of organ damage were observed in the groups that received high doses (3000–5000 mg/kg body weight). In a mouse experiment, Bafor and Igbinuwen [38] demonstrated the safety of a single dose (2.5 to 20 g/kg) of the extract of *F. exasperata* and the risk of prolonged consumption of the plant. Furthermore, Oyetayo *et al.* [39] reported a higher mortality percentage of brine shrimps with increasing concentrations of the extract of *F. exasperata*.

## Conclusion

It can be concluded from the results that *F. exasperata* obtained in Birnin Kebbi is rich in phytochemicals such as flavonoids, saponins, tannins, phenols, cardiac glycosides, and terpenoids. The presence of the mentioned phytochemicals justifies the plant's effectiveness in treating or managing several diseases mentioned in the literature. However, it contains high levels of flavonoids and saponins, suggesting that constantly consuming high doses of the plant for an extended period may induce toxicity. Furthermore, the plant contained non-permissible levels of Cd, Cu, and Pb, demonstrating the plant's potential toxicity. The shrimp lethality assays confirmed these assertions in which low to medium doses (1000 to 2000 ppm) of the extract of the plant tend to be safe, while prolonged high doses (4000 to 8000 ppm) tend to be injurious.

## Recommendations

Based on the findings of this study, prolonged use of high doses of the plant should be avoided.

The plant should be used under the guidance of an expert.

*F. exasperata* can be cultivated or grown at home to enable its monitoring to prevent environmental stress such as contaminants, which may increase its heavy metal and phytochemical content.

A strategy to reduce heavy metals and some phytochemicals in the extracts of the plant to safe doses should be devised or used if it has been devised.

More studies should be carried out to ascertain our claims.

## List of abbreviations

Cu = copper; Pb = lead; Zn = zinc; Cd = cadmium; WHO = World Health Organization; FAO = Food and Agricultural Organization

## Source of funding

This study was supported by a grant from the Tertiary Education Trust Fund (TETFUND), Nigeria (FUBK/2016/BATCH4 RP/2).

## References

[1] Traphagen S. Benefits of Plant-Based Medicine [2017]. Available at https://buffalohealthyliving.com/benefits-plant-based-medicine/ (Accessed Feb 14, 2022).

[2] Yahaya T, Okpuzor J, Ajayi T. Antioxidant Activity of Roselle (*Hibiscus sabdariffa*), Moringa (*Moringaoleifera*), Ginger (*Zingiber officinale*) and 'Ugu' (*Telfairia occidentalis*) in the Lungs of Albino Rats (*Rattus norvegicus*) Exposed to Cement Dust. *Annu Res Rev Biol,* 2014; 4 (5): 736-746. https://doi.org/10.9734/ARRB/2014/5440.

[3] Sánchez M, González-Burgos E, Iglesias I. Current uses and knowledge of medicinal plants in the Autonomous Community of Madrid (Spain): a descriptive cross-sectional study. *BMC Complement Altern Med,* 2020; 20:306. https://doi.org/10.1186/s12906-020-03089-x.

[4] Wahua C, Odogwu BA, Ukomadu J. Macro- and Micro-Morphological Characteristics and Phytochemical Constituents of *Ficus exasperata* Vahl. of Moraceae. *Scholars acad J Biosci*, 2021; 9(4): 111-115. DOI: 10.36347/sajb.2021.v09i04.002.

[5] Anguruwa GT, Oluwadare AO, Fakorede CO, Riki JTB. Soluble extracts present in *Ficus exasperata* (Vahl.) suitable for pulp making. *Proligno*, 2020; 16 (3):36-43. https://www.cabdirect.org/cabdirect/abstract/20203444738.

[6] vanNoort S, Rasplus JY. Figweb: figs and fig wasps of the world [2022]. Available at www.figweb.org (Accessed Feb 14, 2022).

[7] Ashraf K, Haque MR, Amir M, Ahmad N, Ahmad W, Sultan S, et al. An Overview of Phytochemical and Biological Activities: *Ficus deltoidea* Jack and Other *Ficus spp*. *J. Pharm. Bioallied Sci,* 2021; 13(1):11–25. https://doi.org/10.4103/jpbs.JPBS_232_19.

[8] Oyewole OE, Aondoaka TS, Abayomi SJ, Ogundipe TA. Characterization and optimization study of Ficus exasperata extract as corrosion inhibitor for mild steel in seawater. *World Sci News*, 2021; 151: 78-94. bwmeta1.element.psjd-b803fa07-9be4-41ee-83da-6d47b1b9f79c.

[9] Ajala TO, Olusola JA, Odeku OA. Antimicrobial activity of *Ficus exasperata* (Vahl) leaf extract in clinical isolates and its development into herbal tablet dosage form. *J. med. plant. econ. Dev*, 2020; 4(1):a95. https://doi.org/10.4102/jomped.v4i1.95.

[10] Onefeli AO, Fabowale AG. Taxonomic Significance of Leaf Morpho-Anatomical Markers in Identifying *Ficus exasperata* Roxb., *Ficus mucuso* Welw. ex Ficalho and *Ficus thonningii* Blume in Nigeria. *Proceedings*, 2021; 68:1-6. https://doi.org/10.3390/BDEE2021-09482.

[11] Okam PC, Okam CF, Obi E, Unekwe PC. Effect of Sub-chronic Admnistration of Mascum Herbal Pride on Sperm Quality of Male Albino Rats. *Br J Pharm Res*, 2016; 11(4):1-5. DOI: 10.9734/BJPR/2016/21505.

[12] Yahaya T, Shehu K, Isah H, Oladele E, Shemishere U. Toxicological Evaluation of the Leaves of *Guiera senegalensis* (J.F. Gme), *Cassia occidentalis* (Linn), and

*Ziziphus mauritiana* (Lam). *Beni-Suef Univ J Basic Appl Sci*, 2019; 8: 14. https://doi.org/10.1186/s43088-019-0015-y.

[13] Yahaya T, Obaroh IO, Oladele EO. Phytoconstituent Screening of Roselle (*Hibiscus sabdariffa*), Moringa (*Moringa oleifera*), Ginger (*Zingiber officinale*) and Fluted pumpkin (*Telfairia occidentalis*) Leaves. *J. Appl. SCI. Environ,* 2017; 21 (2):253-256. https://doi.org/10.4314/jasem.v21i2.5.

[14] MdAsaduzzaman M, Rana RM, Raqibul H, Hossain MM, Das N. Cytotoxic (Brine Shrimp Lethality Bioassay) and Antioxidant Investigation of Barringtonia Acutangula (L.). *Int J Pharm Sci Res*, 2015; 6 (8): 1179-1184. http://www.ijpsr.info/docs/IJPSR15-06-08-005.pdf.

[15] Yahaya T, Okpuzor J. Variation in Exposure to Cement Dust in Relation to Distance from Cement Company. *Res J Environ Toxicol*, 2011; 5(3): 203-212. http://dx.doi.org/10.3923/rjet.2011.203.212.

[16] Olowa LF, Nuñeza OM. Brine Shrimp Lethality Assay of the Ethanolic Extracts of Three Selected Species of Medicinal Plants from Iligan City, Philippines. *Int Res J Biol Sci*, 2013; 2(11):74-77.

[17] World Health Organization. WHO guidelines for assessing quality of medicinal plant products with reference to contaminants and residues, Geneva, Switzerland, 2007. https://apps.who.int/iris/bitstream/handle/10665/43510/9789241594448_eng.pdf?sequence=1&isAllowed=y.

[18] Samtiya M, Aluko RE, Dhewa T. Plant food anti-nutritional factors and their reduction strategies: an overview. *Food Prod Process and Nutr*, 2020; 2:6. https://doi.org/10.1186/s43014-020-0020-5.

[19] Abraham IG, Ahmad MH. Preliminary sub-acute toxicological assessment of methanol leaves extract of *Culcasia angolensis* (Araceae) in Wistar rats. *Bull Natl Res Cent*, 2021; 45: 226. https://doi.org/10.1186/s42269-021-00686-9.

[20] Lin B, Qi X, Fang L, Zhao L, Zhang R, Jing J, et al. *In vivo* acute toxicity and mutagenic analysis of crude saponins from *Chenopodium quinoa* wild husks. *RSC Adv*, 2021; 11:4829-4841. DOI: 10.1039/D0RA10170B.

[21] Albaayit SFA, Abba Y, Abdullah R, Abdullah N. Evaluation of Antioxidant Activity and Acute Toxicity of *Clausena excavata* Leaves Extract. *Evid.-based Complement. Altern. Med*, 2014; Article ID: 975450. https://doi.org/10.1155/2014/975450.

[22] Kofie W, Osman H, Bekoe SO. Phytochemical Properties of Extracts and Isolated Fractions of Leaves and Stem Bark of *Ficus exasperate*. *World J Pharm Sci*, 2015; 4(12): 91-101. https://www.wjpps.com/Wjpps_controller/abstract_id/4231.

[23] Ajayi OB, Oluyege JO, Olalemi OM, Ilesanmi TM. Nutritional composition, phytochemical screening and antimicrobial properties of the leaf of *Ficus exasperata* (vahl). *Asian J Biol Sci*, 2012; 1(3): 242-246. https://www.ajbls.com/sites/default/files/AsianJBiolLifeSci_1_3_242.pdf.

[24] Nworu PA, Okoye FBC, Akah PA, Esimone CO, Debbab A, Proksch P. Inhibition of inflammatory mediators and reactive oxygen and nitrogen species by some depsidones and diaryl ether derivatives isolated from *Corynespora cassiicola*, an endophytic fungus of *Gongronema latifolium* leaves. *Immunopharmacol Immunotoxicol*, 2013; 35 (6): 662-668. https://doi.org/10.3109/08923973.2013.834930.

[25] Kumar S, Yadav A, Yadav M, Yadav JP. Effect of climate change on phytochemical diversity, total phenolic content and in vitro antioxidant activity of *Aloe vera* (L.) Burm. f. *BMC Res Notes*, 2017; 10(1):60. https://doi.org/10.1186/s13104-017-2385-3.

[26] Adegbaju OD, Otunola GA, Afolayan AJ. Effects of growth stage and seasons on the phytochemical content and antioxidant activities of crude extracts of *Celosia argentea L. Heliyon*, 2020; 6(6): e04086. https://doi.org/10.1016/j.heliyon.2020.e04086.

[27] Rizvi A, Zaidia A, Ameen F, Ahmed B, AlKahtanic MDF, Khana MS. Heavy metal induced stress on wheat: phytotoxicity and microbiological management. *RSC Adv*, 2020; 10:38379-38403. DOI: 10.1039/D0RA05610C.

[28] Yahaya T, Okpuzor J, Ajayi T. The Prophylactic Efficacy of Roselle [*H. sabdariffa*], Moringa [*Moringa oleifera*], Ginger [*Z. officinale*] and Ugwu [*T. occidentalis*] on the Hematology and Serum Protein of Albino Rats [*Rattus norvegicus*] Exposed to Cement Dust. *Res J Med Plants,* 2012; 6:189-196. DOI: 10.3923/rjmp.2012.189.196.

[29] Yahaya T, Ologe O, Yaro C, Abdullahi L, Abubakar H, Gazal J, et al. Quality and Safety Assessment of Water Samples Collected from Wells in Four Emirate Zones of Kebbi State, Nigeria. *Iran J Energy Environ*, 2022; 13(1): 79-86. Doi: 10.5829/ijee.2022.13.01.09.

[30] Rusin M, Domagalska J, Rogala D. Concentration of cadmium and lead in vegetables and fruits. *Sci Rep,* 2021; 11:11913. https://doi.org/10.1038/s41598-021-91554-z.

[31] Lamas GA, Ujueta F, Navas-Acien A. Lead and Cadmium as Cardiovascular Risk Factors: The Burden of Proof Has Been Met. *J Am Heart Assoc*, 2021; 10(10): e018692. https://doi.org/10.1161/JAHA.120.018692.

[32] Bao J, Xing Y, Feng C. Acute and sub-chronic effects of copper on survival, respiratory metabolism, and metal accumulation in *Cambaroides dauricus*. *Sci Rep,* 2020; 10: 16700. https://doi.org/10.1038/s41598-020-73940-1.

[33] Sunmola AF, Ngozi OA, Vivian FD, Christiana FT. Comparative evaluation of the nutritional benefits of some underutilized plants leaves. *J Nat Prod Plant Resour*, 2012; 2(2): 261-266. https://www.scholarsresearchlibrary.com/articles/comparative-evaluation-of-the-nutritional-benefits-of-some-underutilisedplants-leaves.pdf.

[34] Agrahari P, Swati K, Rai S, Singh VK, Singh DK. *Ficus religiosa* Tree Leaves as Bioindicators of Heavy Metals in Gorakhpur City, Uttar Pradesh, India. *Pharmacogn J*, 2018; 10(3): 416-20. DOI: 10.5530/pj.2018.3.68.

[35] Tadesse AW, Gereslassie T, Yan X, Wang J. Determination of Heavy Metal Concentrations and Their Potential Sources in Selected Plants: *Xanthium strumarium* L. (Asteraceae), *Ficus exasperata* Vahl (Moraceae), *Persicaria attenuata* (R.Br) Sojak (Polygonaceae), and *Kanahia laniflora* (Forssk.) R.Br. (Asclepiadaceae) from Awash River Basin, Ethiopia. *Biol Trace Elem Res*, 2019; 191(1): 231–242. https://doi.org/10.1007/s12011-018-1588-3.

[36] Ladipo MK, Doherty FV. Heavy metal analysis and effect of the crude extract of the leaves of Brysocarpus coccineus and Ficus exasperata on some pathogenic organisms. *Int J Biosci*, 2011; 1(2): 17-26. http://citeseerx.ist.psu.edu/viewdoc/download?doi=10.1.1.735.5386&rep=rep1&type=pdf.

[37] Shemishere U, Anyebe D, Yahaya T, Liman U, Bello A. Acute toxicity study of crude methanol leaf extract of *Ficus exasperata* Vahl on male Wistar albino rats. *Biokemistri*, 2020; 32 (1): 47-53.

[38] Bafor EE, Igbinuwen O. Acute toxicity studies of the leaf extract of Ficus exasperata on haematological parameters, body weight and body temperature. *J Ethnopharmacol*, 2009; 123(2): 302–307. https://doi.org/10.1016/j.jep.2009.03.001.

[39] Oyetayo AM, Jose AR, Bada SO, Komolafe TO. Investigation of Phytoconstituents, Antibacterial Activity and Cytotoxic Effect of *Ficus exasperata* Leaf Extracts. *J Adv Microbiol*, 2018; 8(3): 1-8. DOI: 10.9734/JAMB/2018/39733.